\documentclass[conference]{IEEEtran}
\IEEEoverridecommandlockouts
\usepackage{algorithm}                %
\usepackage{algorithmic}              %
\usepackage{amsmath,amssymb,amsfonts} %
\usepackage{cite}                     %
\usepackage{graphicx}                 %
\usepackage{mathtools}                %
\usepackage{mathrsfs}                 %
\usepackage{xcolor}                   %
\def\BibTeX{{\rm B\kern-.05em{\sc i\kern-.025em b}\kern-.08em
    T\kern-.1667em\lower.7ex\hbox{E}\kern-.125emX}}
\ifCLASSOPTIONcompsoc\usepackage[caption=false,font=normalsize,labelfont=sf,textfont=sf]{subfig}\else\usepackage[caption=false,font=footnotesize]{subfig}\fi

\usepackage[bookmarks=false]{hyperref} %

\newcommand{\gauss}[3]{\mathcal{N}(#1; \, #2, \, #3)}

\newcommand\varlist{,\!\makebox[1em][c]{.\hfil.\hfil.},}
\newcommand\prodvarlist{\!\makebox[1em][c]{$\cdot$\hfil$\cdot$\hfil$\cdot$}}

\DeclareMathOperator*{\argmax}{arg\,max}

\newtheorem{theorem}{Theorem}%
\newtheorem{proposition}[theorem]{Proposition}

\begin{document}

\title{%
The Role of Bounded Fields-of-View and Negative Information in Finite Set Statistics (FISST) \\
  \thanks{Keith LeGrand and Silvia Ferrari are with the Laboratory for Intelligent Systems and Controls (LISC), Sibley School of Mechanical and Aerospace Engineering, Cornell University, Ithaca, New York, United States. This work was supported in part by Office of Naval Research Grant N0014-19-1-2266}
}

\author{\IEEEauthorblockN{Keith LeGrand and Silvia Ferrari}
}

\maketitle

\begin{abstract}
  The role of negative information is particularly important to search-detect-track problems in which the number of objects is unknown \emph{a priori}, and the size of the sensor field-of-view is far smaller than that of the region of interest. This paper presents an approach for systematically incorporating knowledge of the field-of-view geometry and position and object inclusion/exclusion evidence into object state densities and random finite set multi-object cardinality distributions.
 The approach is derived for a representative set of multi-object distributions and demonstrated through a sensor planning problem involving a multi-Bernoulli process with up to one-hundred potential targets.
\end{abstract}

\begin{IEEEkeywords}
  Bounded field-of-view, Gaussian mixtures, Gaussian splitting, random finite set theory
\end{IEEEkeywords}

\section{Introduction}
\label{sec:Introductions}

Random finite set (RFS) theory has been proven a highly effective framework for developing and analyzing tracking and sensor planning algorithms in applications involving an unknown number of multiple targets (objects) \cite{MahlerStatisticalMultitargetFusion07, VoLabeledRfsGlmbFilter14, ReuterLmbFilter14, Garcia-fernandezGaussianImplementationPmbm19,HoangSensorManagementMultiBernoulliRenyiMaxCardinalityVariance14,BeardVoidProbabilitiesCauchySchwarzGlmb17,WangMultiSensorControlLmb18}.  To date, however, little attention has been given to the role that bounded FoV and negative information play in the fisst recursive updates for assimilating measurements, or lack thereof, into multi-object probability distributions. Existing algorithms typically terminate object tracks after the object is believed to have left the sensor FoV.  While this approach is suitable when the FoV doubles as the tracking ROI, it is inapplicable when the sensor FoV is much smaller than the ROI and, thus, must be moved or positioned so as to maximize information value \cite{FerrariGeometricDetectIntercept09, WeiGeometricTransversalsSensorPlanning15, GehlySearchDetectTrack18,BuonviriSurveyMultiSensorRfs19}.

Knowledge of object presence inside the FoV is powerful evidence that can be incorporated to update the object pdf in a Bayesian framework.  For example, the absence of detections, referred to as \textit{negative information} may suggest that the object state resides outside the FoV \cite{KochExploitingNegativeEvidence07, SongTargetTrackingStateDependentPd11}.  In contrast, binary-type sensors may indicate that the object is inside the sensor FoV but provide no further localization information.  Particle-based filtering algorithms can accommodate such measurements but require a large number of particles and are computationally expensive.  Another approach \cite{SongTargetTrackingStateDependentPd11} uses GMs to model both the object pdf and the state-dependent probability of detection function.  Though GMs efficiently model some detection probability functions, other simple functions, such as uniform probability over a 3D FoV, require problematically large numbers of components.  Recently, a method based on intermediate particle representation and the EM algorithm has been proposed for forming GMs inside and outside the FoV \cite{WeiDistributedSpaceTrackingFovEm18}.

This paper presents relevant bounded FoV statistics both in the form of state densities and cardinality pmfs.
Section~\ref{sec:GmPartitioning} presents a deterministic method that partitions a GM state density along FoV bounds through recursive Gaussian splitting.  In Section~\ref{sec:FovCardinality}, FoV object cardinality pmfs are derived for some of the most commonly encountered RFS distributions.  Section~\ref{sec:SensorPlacementExample} presents an application of bounded FoV statistics to a sensor placement problem, and conclusions are made in Section~\ref{sec:Conclusions}.

\section{Problem Formulation and Assumptions}
\label{sec:ProblemFormutation}
This paper considers the incorporation of bounded FoV information into algorithms for (multi-)object tracking and sensor planning when the number of objects is unknown and time-varying.
As shown in \cite{BaumgartnerOptimalControlUnderwaterSensor09}, the sensor FoV can be defined as the compact subset $\mathcal{S}(\boldsymbol{q})\subset\mathbb{X}_{\mathrm{p}}$, where $\mathbb{X}_{\mathrm{p}}$ is a subspace of the single-object state space $\mathbb{X}$.  Typically, $\mathbb{X}_{\mathrm{p}}$ represents the object position space and, thus, vectors and densities associated with $\mathbb{X}_{\textrm{p}}$ are referred to as ``position'' quantities in this paper. However, the methods described in the following sections are applicable to any arbitrary subspace of $\mathbb{X}$.  In general, the FoV is a function of the sensor state $\boldsymbol{q}$, which, for example, may consist of the sensor position, orientation, and zoom level.  However, for notational simplicity this dependence is omitted in the remainder of this paper.

Now, let the object state $\boldsymbol{x}$ consist of the kinematic variables that are to be estimated from data via filtering, such as the object position, velocity, turn rate, etc.  Then, the single-object pdf is denoted by $p(\boldsymbol{x})$.  Letting $\boldsymbol{x}_{\mathrm{p}}=\textrm{proj}_{\mathbb{X}_{\mathrm{p}}} \boldsymbol{x}$ denote the state elements that correspond to $\mathbb{X}_{\mathrm{p}}$, an object's presence inside the FoV can be expressed by the generalized indicator function
\begin{align*}
  1_{\mathcal{S}}(\boldsymbol{x}) = \left\{\begin{array}{ll}
      1, & \text{if } \boldsymbol{x}_{\mathrm{p}} \in \mathcal{S} \\
  0, & \text{otherwise}
  \end{array} \right.
\end{align*}
The number of objects and their kinematic states are unknown \emph{a priori}, but can be assumed to consist of discrete and continuous variables, respectively.  The collection of object states is modeled as an RFS $X$ or LRFS $\mathring{X}$, where the single-object labeled state $\mathring{x} = (\boldsymbol{x}, \ell)\in \mathbb{X} \times \mathbb{L}$ consists of a kinematic state vector $\boldsymbol{x}$ and unique discrete label $\ell$.  It is assumed that the prior multi-object distribution is known, e.g., from the output of a multi-object filter, and modeled using either the RFS density $f(X)$ or LRFS density $\mathring{f}(\mathring{X})$.

Throughout this paper, single-object states are represented by lowercase letters~(e.g.~$\boldsymbol{x}$, $\mathring{x}$), while multi-object states are represented by italic uppercase letters~(e.g.~$X$,~$\mathring{X}$).  Bold lowercase letters are used to denote vectors (e.g.~$\boldsymbol{x}$, $\boldsymbol{z}$) and bold uppercase letters are used denote matrices (e.g.~$\boldsymbol{P}$, $\boldsymbol{\Lambda}$).  The accent ``$\mathring{\,\,\,}$'' is used to distinguish labeled states and functions (e.g.~$\mathring{f}$,~$\mathring{x}$,~$\mathring{X}$) from their unlabeled equivalents.  Spaces are represented by blackboard bold symbols~(e.g.~$\mathbb{X}$,~$\mathbb{L}$).

Knowledge of object presence inside the FoV is powerful evidence that can be used to update the object state pdf in a Bayesian framework.  As an example, the single-object state pdf conditioned on its presence inside the FoV can be expressed as
\begin{align}
  \label{eq:1sp}
  p(\boldsymbol{x}\,|\,\mathcal{S}) \propto 1_{\mathcal{S}}(\boldsymbol{x}) p(\boldsymbol{x}) \triangleq p_{\mathcal{S}}(\boldsymbol{x})
\end{align}
Similarly, knowledge of object presence outside of the FoV or equivalently, in the complement set $\mathcal{C}(\mathcal{S}) = \mathbb{X}_{\mathrm{p}} \setminus \mathcal{S}$, can be incorporated, such that
\begin{align}
  \label{eq:1m1sp}
  p(\boldsymbol{x}\,|\,\mathcal{C}(\mathcal{S})) \propto (1 - 1_{\mathcal{S}}(\boldsymbol{x})) p(\boldsymbol{x}) \triangleq p_{\mathcal{C}(\mathcal{S})}(\boldsymbol{x})
\end{align}
Equation~(\ref{eq:1sp}) can be used to model occupancy measurements (e.g.\ the object was detected somewhere in the FoV), and when integrated with respect to $\boldsymbol{x}$, gives the probability that the object is inside the FoV.  Equation~(\ref{eq:1m1sp}) is required to properly incorporate the negative information that an object is not inside $\mathcal{S}$.  For mathematical conciseness, throughout this paper, object \textit{presence} and \textit{absence} are considered rather than object \textit{detection} and \textit{non-detection}. However, Equations~(\ref{eq:1sp})~and~(\ref{eq:1m1sp}) are easily modified to account for detection probabilities.  For example, the event ``object not detected,'' denoted by $\neg D$, is incorporated through the application of Bayes' rule, such that
\begin{align*}
  p(\boldsymbol{x} \,|\, \neg D)
  \propto
  \left(1 - 1_{\mathcal{S}}(\boldsymbol{x})p_{D}(\boldsymbol{x})\right) p(\boldsymbol{x})
\end{align*}
where $p_{D}(\boldsymbol{x})$ is the state-dependent probability of detection.

In RFS-based tracking, single-object densities are, in fact, parameters of the higher-dimensional multi-object density.  Non-Gaussian single-object state densities are often modeled using GMs because they admit closed-form approximations to the multi-object Bayes recursion under certain conditions \cite{VoLabeledRfsGlmbFilter14, VoGaussianMixturePhd06}.  Therefore, in this paper, it is assumed that single-object densities (which are parameters of the higher dimensional multi-object density) are parameterized as
\begin{align*}
  p(\boldsymbol{x}) = \sum_{\ell=1}^{L} w^{(\ell)} \gauss{\boldsymbol{x}}{\boldsymbol{m}^{(\ell)}}{\boldsymbol{P}^{(\ell)}}
\end{align*}
where $L$ is the number of GM components and $w^{(\ell)}$, $\boldsymbol{m}^{(\ell)}$, and $\boldsymbol{P}^{(\ell)}$ are the weight, mean, and covariance matrix of the $\ell$\textsuperscript{th} component, respectively.

The multi-object exponential notation,
\begin{align*}
  h^{A} \triangleq \prod_{a \in A} h(a)
\end{align*}
where $h^{\emptyset}\triangleq 1$, is adopted throughout.
For multivariate functions, the dot $(\cdot)$ denotes the argument of the multi-object exponential, e.g.:
\begin{align*}
  [g(a, \cdot, c)]^{B}
  \triangleq
  \prod_{b \in B}g(a, b, c)
\end{align*}
The exponential notation is used to denote the product space, $\mathbb{X}^{n} = \prod^{n} (\mathbb{X} \times)$.  Exponents of RFSs are used to denote RFSs of a given cardinality, e.g.\ $|X^{n}| = n$, where $n$ is the cardinality.  The operator $\textrm{diag}(\cdot)$ places its input on the diagonal of the zero matrix.  The Kronecker delta function is defined as
\begin{align*}
  \delta_{\boldsymbol{a}}(\boldsymbol{b}) \triangleq
  \left\{\begin{array}{ll}
  1, & \text{if } \boldsymbol{b} = \boldsymbol{a} \\
  0, & \text{otherwise}
  \end{array} \right.
\end{align*}
for any two arbitrary vectors $\boldsymbol{a}, \,\boldsymbol{b} \in \mathbb{R}^{n}$.  The inner product  of two integrable functions $f(\cdot)$ and $g(\cdot)$ is denoted by
\begin{align*}
  \left<f,g\right> = \int f(\boldsymbol{x})g(\boldsymbol{x})\mathrm{d}\boldsymbol{x}
\end{align*}

\section{GM Approximation of FoV-partitioned Densities}
\label{sec:GmPartitioning}
This section presents a method for partitioning the object pdf into truncated densities $p_{\mathcal{S}}(\boldsymbol{x})$ and $p_{\mathcal{C}(\mathcal{S})}(\mathbf{x})$, with supports $\mathbb{X} \setminus \mathcal{C}(\mathcal{S})$ and $\mathbb{X} \setminus \mathcal{S}$, respectively. Focus is given to the single-object state density with the awareness that the method is naturally extended to RFS multi-object densities and algorithms that use GM parameterization. Consider the single-object density $p(\boldsymbol{x})$ parameterized by an $L$-component GM, as follows:
\begin{align*}
  p(\boldsymbol{x})
  =
  p_{\mathcal{S}}(\boldsymbol{x}) + p_{\mathcal{C}(\mathcal{S})}(\boldsymbol{x})
  =
  \sum_{\ell=1}^{L} w^{(\ell)} \mathcal{N}(\boldsymbol{x}; \, \boldsymbol{m}^{(\ell)}, \boldsymbol{P}^{(\ell)})
\end{align*}
One simple approximation of densities partitioned according to the discrete FoV geometry, referred to as FoV-partitioned densities hereon, is found by evaluating the indicator function at the component means \cite{LegrandRelativeSpaceObjectTrackingFusion15}, i.e.:
\begin{align}
  \label{eq:InSTimesPApproxGm}
  p_{\mathcal{S}}(\boldsymbol{x})
  &\approx
  \sum_{\ell=1}^{L} w^{(\ell)}
  1_{\mathcal{S}}(\boldsymbol{m}^{(\ell)})
  \gauss{\boldsymbol{x}}{\boldsymbol{m}^{(\ell)}}{\boldsymbol{P}^{(\ell)}}\\
  \label{eq:NotInSTimesPApproxGm}
  p_{\mathcal{C}(\mathcal{S})}(\boldsymbol{x})
  &\approx
  \sum_{\ell=1}^{L} w^{(\ell)}
  (1 - 1_{\mathcal{S}}(\boldsymbol{m}^{(\ell)}))
  \gauss{\boldsymbol{x}}{\boldsymbol{m}^{(\ell)}}{\boldsymbol{P}^{(\ell)}}
\end{align}
By this approach, components whose means lie inside (outside) the FoV are preserved (pruned), or vice versa.

The accuracy of this mean-based partition approximation depends strongly on the resolution of the GM near the geometric boundaries of the FoV.  Even though the mean of a given component lies inside (outside) the FoV, a considerable proportion of the probability mass may lie outside (inside) the FoV, as is illustrated in Figure~\ref{fig:Example2dSplitOriginalDensity}.  Therefore, the amount of FoV overlap, along with the weight of the component, determines the accuracy of the approximations (Eq.~(\ref{eq:InSTimesPApproxGm})-(\ref{eq:NotInSTimesPApproxGm})).  To that end, the algorithm presented in the following subsection iteratively resolves the GM near FoV bounds by recursively splitting Gaussian components that overlap the FoV bounds.
\newlength\gaussquadwidth
\setlength\gaussquadwidth{4.2cm}
\begin{figure}[h!]
  \centering
  \subfloat[]{\includegraphics[height=\gaussquadwidth]{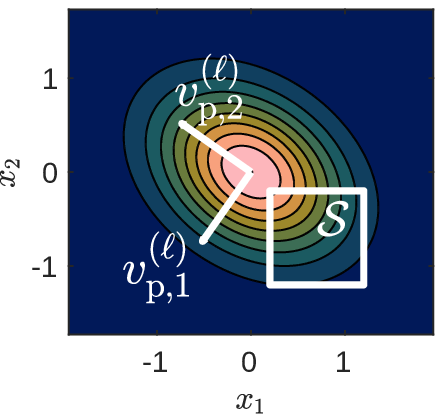}}\hfil
  \subfloat[]{\includegraphics[height=\gaussquadwidth]{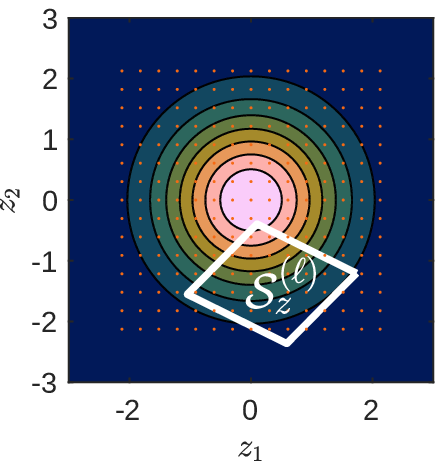}}%
  \caption{Original component density and FoV with covariance eigenvectors overlaid (a), and same component density and FoV after change of variables (b).}
  \label{fig:Example2d}
\end{figure}

\subsection{Gaussian Splitting Algorithm}
The Gaussian splitting algorithm presented in this subsection forms an FoV-partitioned GM approximation of the original GM by using a higher number of components near the FoV boundaries, $\partial \mathcal{S}$, so as to improve the accuracy of the mean-based partition.

Consider for simplicity a two-dimensional example in which the original GM, $p(\boldsymbol{x})$, has a single component whose mean lies outside the FoV, as shown in Figure~\ref{fig:Example2d}a.
The algorithm first applies a change of variables $\boldsymbol{x} \mapsto \boldsymbol{z}$ such that $p(\boldsymbol{z})$ is symmetric, has zero mean and unit variance.
The basis vectors of the space $\mathbb{Z}\ni \mathbf{z}$ correspond to the principal directions of the component's positional covariance.
The same change of variables is applied to the FoV bounds (Figure~\ref{fig:Example2d}b).

A pre-computed point grid is then tested for inclusion in the transformed FoV in order to decide whether to split the component, and if so, along which principal direction.  For each new split component, the process is repeated--if a new component significantly overlaps the FoV boundaries, it may be further split into several smaller components, as illustrated in Figure~\ref{fig:Example2dSplitContours}b.  This process is repeated until stopping criteria are satisfied.  After the GM splitting terminates, $p_{\mathcal{S}}(\boldsymbol{x})$ and $p_{\mathcal{C}(\mathcal{S})}(\boldsymbol{x})$ are approximated by the mean-based partition, as illustrated in Figure~\ref{fig:Example2dSplitInOutFov}.
\setlength\gaussquadwidth{3.7cm}
\begin{figure}[htbp]
  \centering
  \subfloat[]{\includegraphics[height=\gaussquadwidth]{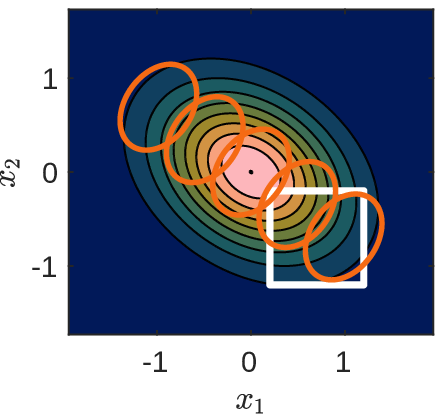}}\hfil
  \subfloat[]{\includegraphics[height=\gaussquadwidth]{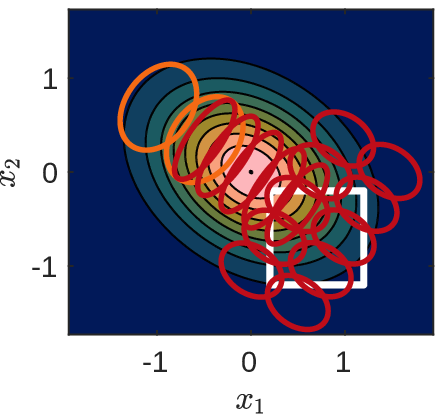}}
  \caption{$1\sigma$ contours of components after first split operation (a), and second split operation (b), where components formed in the second operation are shown in red.}
  \label{fig:Example2dSplitContours}
\end{figure}
\begin{figure}[bhtp]
  \centering
  \subfloat[]{\includegraphics[height=\gaussquadwidth]{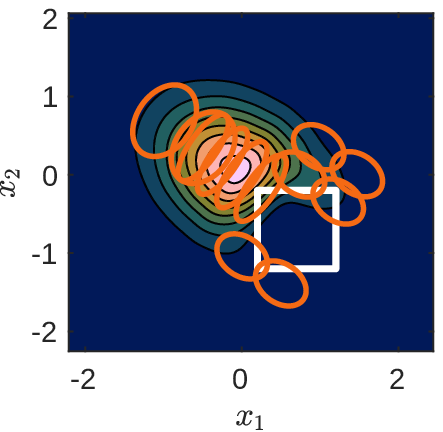}}\hfil
  \subfloat[]{\includegraphics[height=\gaussquadwidth]{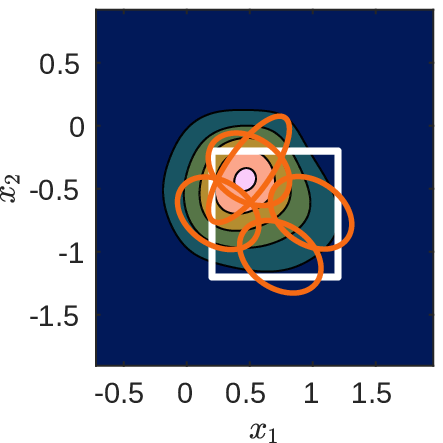}}%
  \caption{The GM approximations to densities $p_{\mathcal{C}(\mathcal{S})}(\boldsymbol{x})$ (a), and $p_{\mathcal{S}}(\boldsymbol{x})$ (b) after two iterations of splitting.}
  \label{fig:Example2dSplitInOutFov}
\end{figure}

\subsection{Univariate Splitting Library}
Splitting is performed efficiently by utilizing a pre-generated library of optimal split parameters for the univariate standard Gaussian $q(x)$, as first proposed in \cite{HuberEntropyApproximationSplitLibrary08} and later generalized in \cite{DemarsEntropyPropagationGaussSplit13}.  The univariate split parameters are retrieved at run-time and applied to arbitrary multivariate Gaussian densities via scaling, shifting, and covariance diagonalization.

Generation of the univariate split library is performed by minimizing the cost function
\begin{align*}
  J = L_{2}(q||\tilde{q}) + \lambda \tilde{\sigma}^{2} \qquad \textrm{s.t.} \sum_{j=1}^{R} \tilde{w}^{(j)}=1
\end{align*}
where
\begin{align*}
  \tilde{q}(x) = \sum \limits_{j=1}^{R} \tilde{w}^{(j)} \gauss{x}{\tilde{m}^{(j)}}{\tilde{\sigma}^{2}}
\end{align*}
for different parameter values $R$, $\lambda$.  The regularization term $\lambda$ balances the importance of using smaller standard deviations $\tilde{\sigma}$ with the minimization of the $L_{2}$ distance. While other distance measures may be used, the $L_{2}$ distance is attractive because it can be computed in closed form for GMs \cite{DemarsEntropyPropagationGaussSplit13}.

\subsection{Change of Variables}
The determination of which components should be split, and if so, along which direction, is simplified by first establishing a change of variables.  For each component with index $\ell$, the change of variables $\boldsymbol{h}^{(\ell)}: \mathbb{X}_{\mathrm{p}} \mapsto \mathbb{Z}$ is applied as follows:
\begin{align}
  \label{eq:TransformationToZ}
  \boldsymbol{z}
  =
  \boldsymbol{h}^{(\ell)}(\boldsymbol{x}_{\mathrm{p}}^{(\ell)}; \boldsymbol{m}_{\mathrm{p}}^{(\ell)}, \boldsymbol{P}_{\mathrm{p}}^{(\ell)})
  \triangleq
  (\boldsymbol{\Lambda}_{\mathrm{p}}^{(\ell)})^{-\frac{1}{2}}\boldsymbol{V}_{\mathrm{p}}^{(\ell)T} (\boldsymbol{x}_{\mathrm{p}} - \boldsymbol{m}_{\mathrm{p}}^{(\ell)})
\end{align}
where
\begin{align*}
  \boldsymbol{V}_{\mathrm{p}}^{(\ell)}
  &=
  [\boldsymbol{v}_{\mathrm{p},1}^{(\ell)} \quad \cdots \quad \boldsymbol{v}_{\mathrm{p},n_{\mathrm{p}}}^{(\ell)}]\nonumber\\
  (\boldsymbol{\Lambda}_{\mathrm{p}}^{(\ell)})^{-1/2}
  &=
  \textrm{diag}
  \left(
    \left[
      \tfrac{1}{\sqrt{\lambda_{\mathrm{p}, 1}^{(\ell)}}}\quad  \cdots \quad \tfrac{1}{\sqrt{\lambda_{\mathrm{p},n_{\mathrm{p}}}^{(\ell)}}}
    \right]
  \right)\nonumber
\end{align*}
and $\boldsymbol{m}_{\mathrm{p}}^{(\ell)}$ is the $n_{\mathrm{p}}$-element position portion of the full-state mean, and the columns of $\boldsymbol{V}_{\mathrm{p}}^{(\ell)}$ are the normalized eigenvectors of the position-marginal covariance $\boldsymbol{P}_{\mathrm{p}}^{(\ell)}$, with $\boldsymbol{v}_{\mathrm{p}, i}^{(\ell)}$ corresponding to the $i$\textsuperscript{th} eigenvalue $\lambda_{\mathrm{p}, i}^{(\ell)}$.  In the transformed space,
\begin{align*}
  p_{z}(\boldsymbol{z}) = \gauss{\boldsymbol{z}}{\boldsymbol{0}}{\boldsymbol{I}}
\end{align*}
Note that, in defining the transformation over $\mathbb{X}_{\mathrm{p}}$, the same transformation can be applied to the FoV, such that
\begin{align}
  \label{eq:Sz}
  \mathcal{S}_{z}^{(\ell)} = \{\boldsymbol{h}^{(\ell)}(\boldsymbol{x}_{\mathrm{p}}; \boldsymbol{m}_{\mathrm{p}}^{(\ell)}, \boldsymbol{P}_{\mathrm{p}}^{(\ell)}) : \boldsymbol{x}_{\mathrm{p}} \in \mathcal{S}\}
\end{align}

In $\mathbb{Z}$, the Euclidean distances to boundary points of $\mathcal{S}_{z}^{(\ell)}$ can be interpreted as probabilistically normalized distances.  In fact, the Euclidean distance of a point $\boldsymbol{z}$ from the origin in $\mathbb{Z}$ corresponds exactly to the Mahalanobis distance between the corresponding point $\boldsymbol{x}_{\mathrm{p}}$ and the original position-marginal component.

\subsection{Component Selection and Collocation Points}
Components are selected for splitting if they have sufficient weight and significant statistical overlap of the FoV boundaries ($\partial \mathcal{S}$).  For components of sufficient weight, the change of variables is applied to the FoV to obtain $\mathcal{S}^{(\ell)}_{z}$ per Equation~(\ref{eq:Sz}).  The overlap of the original component on $\mathcal{S}$ is then equivalent to the overlap of the standard Gaussian distribution on $\mathcal{S}^{(\ell)}_{z}$, which is quantified using a grid of collocation points on $\mathbb{Z}$.  Define a uniform grid of collocation points $\{\bar{\boldsymbol{z}}_{i_{1}, \dots i_{n_{\mathrm{p}}}}\}$ on $\mathbb{Z}$ such that
\begin{align*}
  \bar{\boldsymbol{z}}_{i_{1}, \dots, i_{n_{\mathrm{p}}}}
  &=
  [\bar{z}_{1}(i_{1}) \, \dots \bar{z}_{n_{\mathrm{p}}}(i_{n_{\mathrm{p}}})]^{T}\\
  \bar{z}_{j}(i_{j})
  &=
  -\zeta + 2\zeta \left(\frac{i_{j} - 1}{N-1}\right) , \quad i_{j} = 1 \varlist N
\end{align*}
where $\zeta$ is a user-specified bound for the grid and $N$ is the number of points per dimension.  An inclusion variable is defined as
\begin{align*}
  d_{i_{1},\dots, i_{n_{\mathrm{p}}}}^{(\ell)}
  \triangleq
  1_{\mathcal{S}_{z}^{(\ell)}}(\bar{\boldsymbol{z}}_{i_{1}, \dots, i_{n_{\mathrm{p}}}})
\end{align*}
A function $s_{\mathcal{S}_{z}}^{(\ell)}(\cdot)$ is established to mark total inclusion or total exclusion as
\begin{align*}
  s_{\mathcal{S}_{z}}^{{(\ell)}}(\bar{\boldsymbol{z}}_{i_{1}, \dots, i_{n_{\mathrm{p}}}})= \prod_{i_{1}, \dots, i_{n_{\mathrm{p}}}} \delta_{d_{1, \dots, 1}^{(\ell)}}
  (d_{i_{1}, \dots, i_{n_{\mathrm{p}}}}^{(\ell)})
\end{align*}
which is equal to unity if all grid points lie inside of $\mathcal{S}_{z}^{(\ell)}$ or all grid points lie outside of $\mathcal{S}_{z}^{(\ell)}$, and is zero otherwise.  If either all or no points are included, no splitting is required.  Otherwise, the component is marked for splitting.

\subsection{Positional Split Direction}
Rather than split a component along each of its principal directions, a more judicious selection can be made by limiting split operations to a single direction (per component) per recursion.  Thus, by performing one split per component per recursion, the component selection criteria are re-evaluated, reducing the overall number of components generated.  In the aforementioned two-dimensional example, only a subset of new components generated from the first split are selected for further splitting as shown in Figure~\ref{fig:Example2dSplitContours}b.

The split direction is chosen based on the relative geometry of the FoV, and thus positional vectors are of interest. Choosing the best positional split direction is a challenging problem.  Ideally, splitting along the chosen direction should minimize the number of splits required in the next iteration as well as improve the accuracy of the partition approximation applied after the final iteration.  The computational complexity of exhaustive optimization of the split direction would likely negate the computational efficiency of the overall algorithm.  Instead, to minimize the number of splits required in the next iteration, the positional split direction is chosen as the direction that is orthogonal to the most grid planes of consistent inclusion/exclusion.  The plane of constant $z_{j}=\bar{z}_{j}(i_{j})$ is consistently inside or consistently outside if
\begin{align*}
  s_{j}^{(\ell)}(i_{j}) = \prod_{i_{1},\dots,i_{j-1}, i_{j+1}, i_{n_{\mathrm{p}}}} \delta_{d_{1,\dots, i_{j}, \dots, 1}^{(\ell)}}
  (d_{i_{1},\dots, i_{j}, \dots, i_{n_{\mathrm{p}}}}^{(\ell)})
\end{align*}
is equal to unity.  The optimal positional split direction is then given by the eigenvector $\boldsymbol{v}_{\mathrm{p}, j^{*}}$, where the optimal eigenvector index is found as
\begin{align}
  \label{eq:BestPositionSplitDirection}
  j^{*} = \argmax_{j} \left(\sum_{i_{j}} s_{j}^{(\ell)}(i_{j})\right)
\end{align}
For notational simplicity, the implicit dependence of $j^{*}$ on the component index $\ell$ is omitted.
For example, referring back to the two-dimensional example and Figure~\ref{fig:Example2d}b, there are more rows than columns that are consistently inside or outside the transformed FoV, and thus $j^{*}=2$ is chosen as the desired positional split direction index.  In the case where multiple maxima exist, the eigenvector with largest eigenvalue is selected, which corresponds to the direction of largest variance among the maximizing eigenvectors.

\subsection{Multivariate Split of Full-state Component}
Gaussian splitting must be performed along the principal directions of the full-state covariance.  The general multivariate split approximation, splitting along the $k$\textsuperscript{th} eigenvector $\boldsymbol{v}_{k}^{(\ell)}$ is given by \cite{DemarsEntropyPropagationGaussSplit13}
\begin{align}
  \label{eq:SplitMultivariate}
  w^{(\ell)} \gauss{\boldsymbol{x}}{\boldsymbol{m}^{(\ell)}}{\boldsymbol{P}^{(\ell)}} \approx \sum_{j=1}^{R} w^{(\ell,j)} \gauss{\boldsymbol{x}}{\boldsymbol{m}^{(\ell,j)}}{\boldsymbol{P}^{(\ell,j)}}
\end{align}
where
\begin{gather*}
  w^{(\ell,j)}
  =
  \tilde{w}^{(j)} w^{(\ell)},
  \quad
  \boldsymbol{m}^{(\ell, j)} = \boldsymbol{m}^{(\ell)} + \sqrt{\lambda_{k}^{(\ell)}}
  \tilde{m}^{(j)} \boldsymbol{v}_{k}^{(\ell)}\\
  \boldsymbol{P}^{(\ell,j)}
  \!=\!
  \boldsymbol{V}^{(\ell)} \boldsymbol{\Lambda}^{(\ell)} \boldsymbol{V}^{(\ell)T},
  \quad
  \boldsymbol{\Lambda}^{(\ell)}
  \!=\!
  \textrm{diag}
  \left([
    \lambda_{1} \, \cdots \, \tilde{\sigma}^{2} \lambda_{k} \, \cdots \, \lambda_{n}
  ]\right)
\end{gather*}
and the optimal univariate split parameters $\tilde{w}^{(j)}$, $\tilde{m}^{(j)}$, and $\tilde{\sigma}$ are found from the pre-computed split library given the number of split components $R$ and regularization parameter $\lambda$.  In general, the positional components of the full-state eigenvectors will not perfectly match the desired positional split vector due to correlations between the states.  Rather, the actual full-state split is performed along $\boldsymbol{v}_{k*}^{(\ell)}$, where the optimal eigenvector index is found according to
\begin{align}
  \label{eq:BestFullStateSplitDirection}
  k^{*} = \argmax_{k} \big|\big[\boldsymbol{v}_{\mathrm{p},j^{*}}^{(\ell)T} \,\, \boldsymbol{0}^{T}\big] \boldsymbol{v}^{(\ell)}_{k}\big|
\end{align}
where, without loss of generality, a specific state convention is assumed such that position states are first in element order.
\subsection{Recursion and Role of Negative Information}
The splitting procedure is applied recursively, as detailed in Algorithm~\ref{alg:SplitForFov}.  The recursion is terminated when no remaining components satisfy the criteria for splitting.  Each recursion further refines the GM near the FoV bounds to improve the approximations of Equations~(\ref{eq:InSTimesPApproxGm})-(\ref{eq:NotInSTimesPApproxGm}).  However, because a Gaussian component's split approximation (Eq.\ \ref{eq:SplitMultivariate}) does not perfectly replicate the original component, a small error is induced with each split.  Given enough recursions, this error may become dominant.  In the authors' experience, the recursion is terminated well before the cumulative split approximation error dominates.

\begin{algorithm}
\begin{algorithmic}
  \caption{\texttt{split\_for\_fov(}$\{w^{(\ell)}, \boldsymbol{m}^{(\ell)}, \boldsymbol{P}^{(\ell)}\}_{\ell=1}^{L}$, $w_{\min}$, $\mathcal{S}$, $R$, $\lambda$\texttt{)}}
  \label{alg:SplitForFov}
  \STATE \texttt{split} $\leftarrow \{\}$, \texttt{no\_split} $\leftarrow \{\}$
  \IF {$L=0$}
    \RETURN \texttt{split}
  \ENDIF
  \FOR {$\ell=1,\dots,L$}
    \IF {$w^{(\ell)}<w_{\min}$}
      \STATE add $\{w^{(\ell)}, \boldsymbol{m}^{(\ell)}, \boldsymbol{P}^{(\ell)}\}$ to \texttt{no\_split}
      \STATE \textbf{continue}
    \ENDIF
    \STATE Compute $\mathcal{S}^{(\ell)}_{z}$ accrd.\ to Eq.~(\ref{eq:Sz})
    \IF {$s_{\mathcal{S}_{z}^{(\ell)}}(\bar{\boldsymbol{z}}_{i_{1}, \dots, i_{n_{\mathrm{p}}}})=1$}
    \STATE add $\{w^{(\ell)}, \boldsymbol{m}^{(\ell)}, \boldsymbol{P}^{(\ell)}\}$ to \texttt{no\_split}
    \ELSE
    \STATE $j^{*} \gets $ Eq.~(\ref{eq:BestPositionSplitDirection}) , $k^{*} \gets $ Eq.~(\ref{eq:BestFullStateSplitDirection})
    \STATE $\{w^{(\ell,j)}, \boldsymbol{m}^{(\ell,j)}, \boldsymbol{P}^{(\ell,j)}\}_{j=1}^{R} \gets$ Eq.~(\ref{eq:SplitMultivariate}) with $k=k^{*}$
    \STATE add $\{w^{(\ell,j)}, \boldsymbol{m}^{(\ell,j)}, \boldsymbol{P}^{(\ell,j)}\}_{j=1}^{R}$ to \texttt{split} \ENDIF
  \ENDFOR
  \STATE \texttt{split}$\leftarrow$\texttt{split\_for\_fov(split}, $w_{\min}$, $\mathcal{S}$, $R$, $\lambda${)}
  \RETURN \texttt{split} $\cup$ \texttt{no\_split}
\end{algorithmic}
\end{algorithm}

One of the many potential applications of the recursive algorithm presented in this section involves incorporating the evidence of non-detections, or negative information, in single- or multi-object filtering. To demonstrate, a single-object filtering problem with a bounded square FoV is considered where, in three subsequent sensor reports, no object is detected. The true object position and constant velocity are unknown but are distributed according to a known GM pdf at the first time step. As the initial pdf is propagated over time, the position-marginal pdf travels from left to right, as pictured in Figure~\ref{fig:GaussSplitRectFov}. For simplicity, the probability of detection inside the FoV is assumed equal to one. At each time step, the GM is refined by Algorithm \ref{alg:SplitForFov} using $w_{\min}=0.01$, $R=3$, and $\lambda=0.001$. Then, the mean-based partition approximation is applied (Eq.~\ref{eq:NotInSTimesPApproxGm}) and the updated filtering density is found (Eq.~\ref{eq:1m1sp}).  By this approach, the number of components may increase over time but can be reduced as needed through component merging and pruning.

\newlength\gausssplitwidth
\setlength\gausssplitwidth{2.85cm}
\begin{figure}[htbp]
  \centering
  \includegraphics[width=\gausssplitwidth]{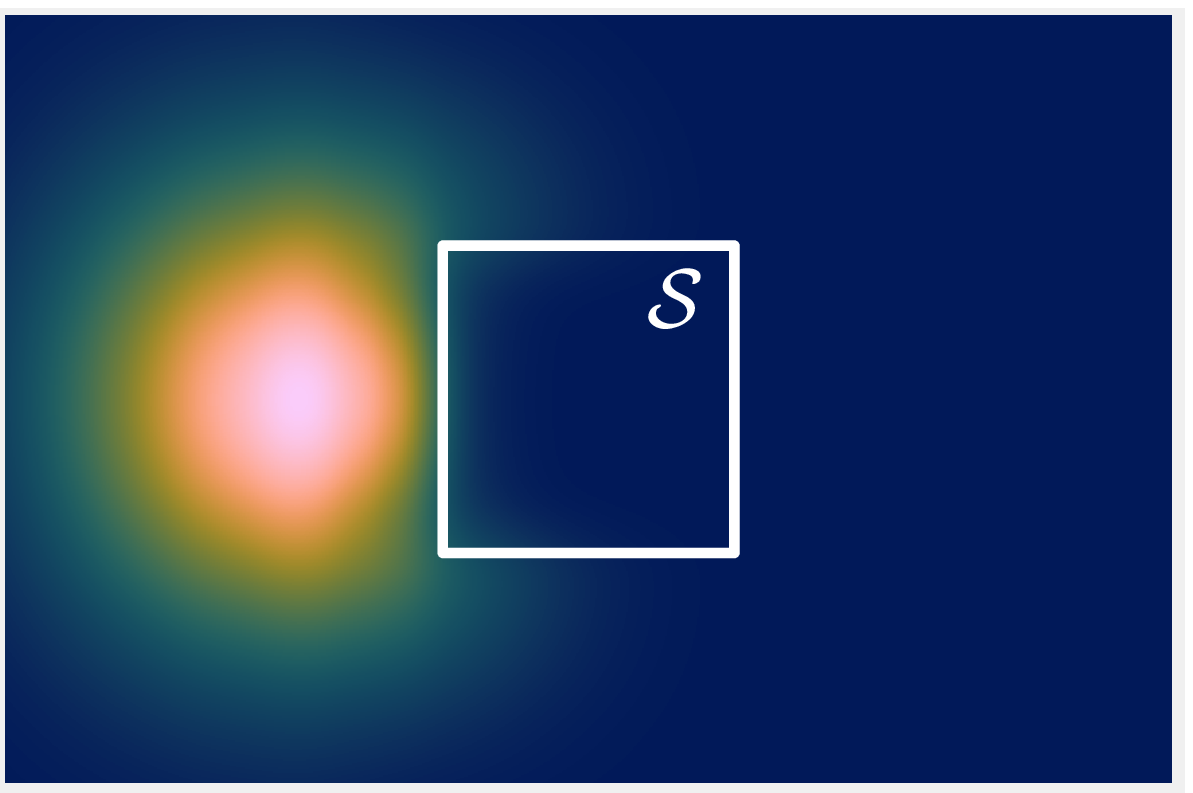}
  \includegraphics[width=\gausssplitwidth]{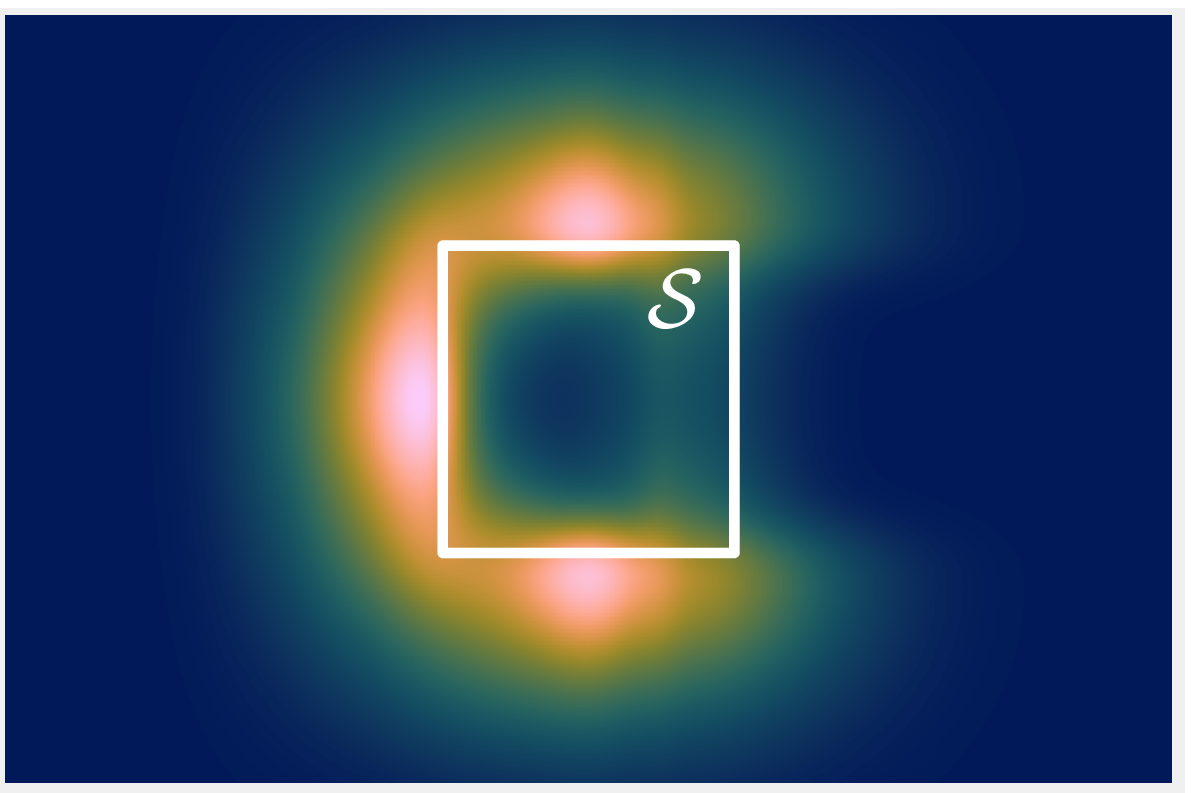}
  \includegraphics[width=\gausssplitwidth]{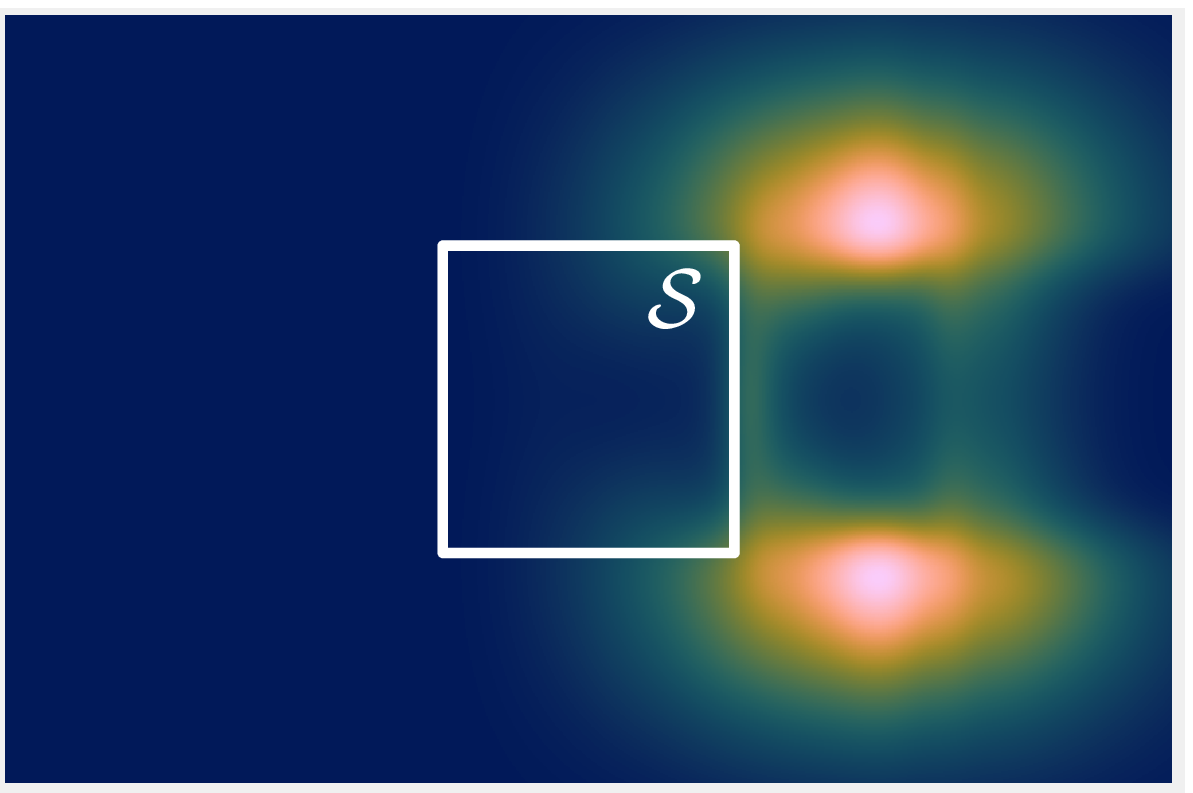}
  \caption{Negative information, comprised of absence of detections inside the sensor FoV $\mathcal{S}$, is used to update the object pdf as the object moves across the ROI.}
  \label{fig:GaussSplitRectFov}
\end{figure}

\section{FoV Cardinality Distribution}
\label{sec:FovCardinality}
This section presents pmfs for the cardinality of objects inside a bounded FoV $\mathcal{S}$ given different multi-object workspace densities $f(\cdot)$.  The Poisson, IIDC, MB, and GLMB distributions are considered in Subsections~\ref{ss:poisson},~\ref{ss:iidc},~\ref{ss:mb}, and~\ref{ss:glmb}, respectively.

The probability of $n$ objects existing inside FoV $\mathcal{S}$ conditioned on $X$ can be written in terms of the indicator function as
\begin{align}
  \rho_{\mathcal{S}}(n \,|\, X) = \sum_{X^{n} \subseteq X} [1_{\mathcal{S}}(\cdot)]^{X^{n}} [1 - 1_{\mathcal{S}}(\cdot)]^{X \setminus X^{n}}
  \label{eq:ConditionalCardinality}
\end{align}
where the summation is taken over all subsets $X^{n}\subseteq X$ with cardinality $n$.  Given the RFS density $f(X)$, the FoV cardinality distribution is obtained via the set integral as
\begin{align*}
  \rho_{\mathcal{S}}(n) = \int \rho_{\mathcal{S}}(n\,|\,X) f(X) \delta X
\end{align*}
Expanding the integral,
\begin{align}
  \label{eq:FovCardinalityGeneralExpanded}
  &\rho_{\mathcal{S}}(n) =  \\
  &\quad\sum_{m=n}^{\infty} \frac{1}{m!}
  \int\limits_{\mathbb{X}^{m}} \rho_{\mathcal{S}}(n\,|\,\{\boldsymbol{x}_{1} \varlist \boldsymbol{x}_{m}\}) f(\{\boldsymbol{x}_{1} \varlist \boldsymbol{x}_{m}\}) \mathrm{d} \boldsymbol{x}_{1} \prodvarlist \mathrm{d} \boldsymbol{x}_{m} \nonumber
\end{align}
\textit{Remark}: The results presented in this section can be trivially extended to express the predicted cardinality of object-originated \textit{detections} $Z$ (excluding false alarms) by noting that
\begin{align*}
  \rho_{\mathcal{S}}(n_{Z}\,|\, X) = \sum_{X^{n} \subseteq X} [p_{D}(\cdot) 1_{\mathcal{S}}(\cdot)]^{X^{n}} [1 - p_{D}(\cdot)1_{\mathcal{S}}(\cdot)]^{X \setminus X^{n}}
\end{align*}
where $n_{Z}=|Z|$.

\subsection{Poisson Distribution}
\label{ss:poisson}
The density of a Poisson-distributed RFS is
\begin{align}
  f(X) = e^{-N_{X}} [D]^{X}
  \label{eq:PoissonDensity}
\end{align}
where $N_{X}$ is the global cardinality mean, and $D(\boldsymbol{x})$ is the PHD, or intensity function, of $X$, which is defined on the single-object space $\mathbb{X}$.  One important property of the PHD is that its integral over a closed set on $\mathbb{X}$ yields the expected number of objects within that set, i.e.
\begin{align}
  \label{eq:PoissonIntegralIsExpectedCardinality}
  E[|X\cap T|] = \int_{T} D(\boldsymbol{x}) \mathrm{d} \boldsymbol{x}
\end{align}
\begin{proposition}
  Given a Poisson-distributed RFS with PHD $D(\boldsymbol{x})$ and global cardinality mean $N_{X}$, the cardinality of objects inside the field of view $\mathcal{S}\subseteq \mathbb{X}$ is distributed according to
  \begin{align}
    \rho_{\mathcal{S}}(n) = \sum_{m=n}^{\infty}\frac{e^{-N_{X}}}{n!(m-n)!}
    \left<
    1_{\mathcal{S}}, D
    \right>^{n}
    \left<
    1-1_{\mathcal{S}}, D
    \right>^{m-n}
    \label{eq:PoissonFovCardinality}
  \end{align}
  \label{pro:PoissonFovCardinality}
\end{proposition}
\textit{Proof:}
Substituting Equation~(\ref{eq:PoissonDensity}) into Equation~(\ref{eq:FovCardinalityGeneralExpanded}),
\begin{align}
  \rho_{\mathcal{S}}(n)
  &=
  \sum_{m=n}^{\infty}
  \frac{1}{m!}
  e^{-N_{X}}
  \int_{\mathbb{X}^{m}} \sum_{X^{n}\subseteq X}
  [1_{\mathcal{S}}(\cdot) D(\cdot)]^{X^{n}} \nonumber \\
  &\qquad \cdot
  [(1 - 1_{\mathcal{S}}(\cdot)) D(\cdot)]^{X \setminus X^{n}}
  \mathrm{d} \boldsymbol{x}_{1} \cdots \mathrm{d} \boldsymbol{x}_{m}
  \label{eq:PoissonFovCardinalityXmIntegral}
\end{align}
The nested integrals of Equation~(\ref{eq:PoissonFovCardinalityXmIntegral}) can be distributed, rewriting the second sum over $n$-cardinality index sets $\mathcal{I}^{n}$ as
\begin{align*}
  \rho_{\mathcal{S}}(n)
  &=
  \sum_{m=n}^{\infty}
  \frac{1}{m!}
  e^{-N_{X}}
  \, \,
  \sum_{\mathclap{\mathcal{I}^{n}\subseteq\{1..m\}}} \quad
  \left[
    \int 1_{\mathcal{S}}(\boldsymbol{x}_{(\cdot)}) D(\boldsymbol{x}_{(\cdot)})\mathrm{d} \boldsymbol{x}_{(\cdot)}
  \right]^{\mathcal{I}^{n}} \nonumber \\
  &\qquad \cdot
  \left[
    \int(1 - 1_{\mathcal{S}}(\boldsymbol{x}_{(\cdot)})) D(\boldsymbol{x}_{(\cdot)})
  \right]^{\{1..m\} \setminus \mathcal{I}^{n}}
\end{align*}
where the shorthand $\{1..m\}$ is used to denote the set of integers $\{1 \varlist m\}$.  Note that the value of the integrals is independent of the variable index, and thus
\begin{align*}
  \rho_{\mathcal{S}}(n)
  &=
  \sum_{m=n}^{\infty}
  e^{-N_{X}}
  \frac{1}{m!}
  \frac{m!}{n!(m-n)!}
  \left<
    1_{\mathcal{S}}, D
  \right>^{n}
  \left<
    1 - 1_{\mathcal{S}}, D
  \right>^{m-n}
\end{align*}
from which Equation~(\ref{eq:PoissonFovCardinality}) follows.
\null\hfill$\square$

\textit{Remark}: Computation of Equation~(\ref{eq:PoissonFovCardinality}) requires only one integral computation; namely $\big<1_{\mathcal{S}}, D\big>$, which can be found either by summing the weights of Equation~(\ref{eq:InSTimesPApproxGm}) or through Monte Carlo integration.
Using the integral property of the PHD (Eq.~\ref{eq:PoissonIntegralIsExpectedCardinality}), the integral
\begin{align*}
  \big<1 - 1_{\mathcal{S}}, D\big>=N_{X} - \big<1_{\mathcal{S}}, D\big>
\end{align*}
Furthermore, for $m\gg N_{X}$, the summand of Equation~(\ref{eq:PoissonFovCardinality}) is negligible, and the infinite sum can be safely truncated at an appropriately chosen $m=m_{\max}(N_{X})$.

\subsection{Independent Identically Distributed Cluster Distribution}
\label{ss:iidc}

The density of an IIDC RFS is
\begin{align}
  f(X)
  =
  |X|!
  \cdot
  \rho(|X|)
  [p]^{X} \,,
  \label{eq:IidcDensity}
\end{align}
where $\rho(n)$ is the cardinality pmf and $p(\boldsymbol{x})$ is the single-object state pdf.

\begin{proposition}
  Given an IIDC-distributed RFS with cardinality pmf $\rho(\cdot)$ and state density $p(\cdot)$, the cardinality of objects inside the FoV $\mathcal{S}$ is distributed according to
  \begin{align}
    \rho_{\mathcal{S}}(n)
    =
    \sum_{m=n}^{\infty} \rho (m)
    \binom{m}{n}
    \big< 1_{\mathcal{S}}, p\big>^{n}
    \big< 1 - 1_{\mathcal{S}}, p\big>^{m-n}
    \label{eq:IidcFovCardinality}
  \end{align}
  where $\binom{m}{n}$ is the binomial coefficient.
  \label{pro:IidcFovCardinality}
\end{proposition}

\textit{Proof}:
Substituting Equation~(\ref{eq:IidcDensity}) into Equation~(\ref{eq:FovCardinalityGeneralExpanded}),
\begin{align*}
  &\rho_{\mathcal{S}}(n) = \sum\limits_{m=n}^{\infty} \frac{1}{m!} m! \rho(m)\\
  &\quad \int\limits_{\mathbb{X}^{m}}
  \sum_{X^{n}\subseteq X}
  \cdot
  [1_{s}(\cdot) p(\cdot)]^{X^{n}}
  [(1 - 1_{s}(\cdot)) p(\cdot)]^{X \setminus X^{n}}
  \mathrm{d} \boldsymbol{x}_{1} \prodvarlist \mathrm{d} \boldsymbol{x}_{m}
\end{align*}
The integral can be moved inside the products so that
\begin{align}
  \rho_{\mathcal{S}}(n)
  =
  \sum\limits_{m=n}^{\infty} \rho(m)
  \sum_{\mathcal{I}^{n}\subseteq \{1..m\}}
  \left[
    \int 1_{s}(\boldsymbol{x}_{(\cdot)}) p(\boldsymbol{x}_{(\cdot)}) \mathrm{d} \boldsymbol{x}_{(\cdot)}
  \right]^{\mathcal{I}^{n}} \nonumber\\
  \cdot
  \left[
    \int (1 - 1_{s}(\boldsymbol{x}_{(\cdot)})) p(\boldsymbol{x}_{(\cdot)}) \mathrm{d} \boldsymbol{x}_{(\cdot)}
  \right]^{\{1..m\} \setminus \mathcal{I}^{n}}
  \label{eq:IidcFovCardinalityIndexSet}
\end{align}
Equation~(\ref{eq:IidcFovCardinality}) follows from Equation~(\ref{eq:IidcFovCardinalityIndexSet}) by noting that there are $\binom{m}{n}$ unique unordered $n$-cardinality index subsets of $\{1 \varlist m\}$.
\null\hfill$\square$

\subsection{Multi-Bernoulli Distribution}
\label{ss:mb}
The density of a MB distribution is \cite[p.\ 102]{MahlerAdvancesStatisticalMultitargetFusion14}
\begin{align}
  f(X)
  =
  \left[
    \left(
      1 - r^{(\cdot)}
    \right)
  \right]^{\{1.. M\}}
  \sum\limits_{\mathclap{\vphantom{\big[}1 \leq i_{1} \neq \cdots \neq i_{n} \leq M}}
    \quad
  \left[
    \dfrac{
      r^{i_{(\cdot)}} p^{i_{(\cdot)}}(\boldsymbol{x}_{(\cdot)})
    }{
      1 - r^{i_{(\cdot)}}
    }
  \right]^{\mathrlap{\{1..n\}}}
  \label{eq:MbDensity}
\end{align}
where $M$ is the number of MB components and maximum possible object cardinality, $r^{i}$ is the probability that the $i$\textsuperscript{th} object exists, and $p^{i}(\boldsymbol{x})$ is the single-object state density of the $i$\textsuperscript{th} object if it exists.

\begin{proposition}
\label{pro:MbFovCardinality}%
  Given at MB density of the form of Equation~(\ref{eq:MbDensity}), the cardinality of objects inside the FoV $\mathcal{S}$ is distributed according to
  \begin{align}
    &\rho_{\mathcal{S}}(n) =
    \left[
      \left(
        1 - r^{(\cdot)}
      \right)
    \right]^{\{1 .. M\}} \nonumber \\
    &\quad
    \cdot
    \sum\limits_{\mathclap{\mathcal{I}_{1} \uplus \mathcal{I}_{2}\uplus \mathcal{I}_{3}}}
    \delta_{n}(|\mathcal{I}_{1}|)
    \left[
      \dfrac{%
        \left< 1_{\mathcal{S}}, r^{(\cdot)} p^{(\cdot)}\right>
      }{%
        1 - r^{(\cdot)}
      }
    \right]^{\mathcal{I}_{1}}
    \left[
      \dfrac{%
        \left<1 - 1_{\mathcal{S}},
        r^{(\cdot)} p^{(\cdot)}\right>
      }{%
        1 - r^{(\cdot)}
      }
    \right]^{\mathcal{I}_{2}}
    \label{eq:MbFovCardinality}
  \end{align}
where the summation is taken over all mutually exclusive index partitions $\mathcal{I}_{1}\uplus \mathcal{I}_{2}\uplus \mathcal{I}_{3}=\{1..M\}$.
\end{proposition}

Proof of Proposition~\ref{pro:MbFovCardinality} is given in Appendix~\ref{a:ProofOfMbFovCardinality}.
Following the same procedure, similar results for the LMB \cite{ReuterLmbFilter14} and MBM \cite{WilliamsMarginalMultiBernoulliMixturePGFL15} RFS distributions may be obtained.

Direct computation of Equation~(\ref{eq:MbFovCardinality}) is only feasible for small $M$ due to the sum over all permutations $\mathcal{I}_{1} \uplus \mathcal{I}_{2} \uplus \mathcal{I}_{3}$.  For large $M$, a stochastic approximation may be used, as detailed in Algorithm~2 and summarized as follows.  For each MB component, the integral $\left<1_{\mathcal{S}}, p^{(i)}\right>$ is computed either by summing the weights of the partitioned GM or by Monte Carlo integration.  Using the integral results, the probability of object $i$ existing inside the FoV is found as
\begin{align*}
  r_{\mathcal{S}}^{(i)} = r^{(i)}\left< 1_{\mathcal{S}},p^{(i)}\right>
\end{align*}
These probabilities are then sampled in $N_s$ Monte Carlo trials to randomly generate $\bar{n}_{i,j}$ which is unity if the object $i$ is in $\mathcal{S}$ in the $j$\textsuperscript{th} trial and zero otherwise.  The cardinality of each random trial is tallied, and the probability of $n$ objects existing inside the FoV is given by the proportion of the number of trials with $n$ objects with respect to the total number of trials.
\begin{algorithm}[htbp]
\begin{algorithmic}
  \caption{Stochastic MB FoV Cardinality}
  \label{alg:StochasticMbCardinality}
  \FOR{$i=1,\dots,M$}
    \STATE $r_{\mathcal{S}}^{(i)} \gets r^{(i)}\left<1_{\mathcal{S}}, p^{(i)}\right>$
  \ENDFOR
  \FOR{$j=1,\dots,N_{s}$}
    \FOR{$i=1,\dots,M$}
    \STATE $u \sim \textrm{Uniform}[0,1]$
    \STATE $\bar{n}_{i,j} \gets r_{\mathcal{S}}^{(i)} \geq u$
    \ENDFOR
    \STATE $\bar{n}_{j} \gets \sum_{i=1}^{M} \bar{n}_{i,j}$
  \ENDFOR
  \STATE $\rho_{\mathcal{S}}(n) \gets \frac{1}{N_{s}}\sum_{j=1}^{N_{s}} \delta_{n}(\hat{n}_{j})$
\end{algorithmic}
\label{alg:mbFovCardinality}
\end{algorithm}

\subsection{Generalized Labeled Multi-Bernoulli Distribution}
\label{ss:glmb}
The density of a GLMB distribution is given by \cite{VoLabeledRfsGlmbFilter14}
\begin{align}
  \label{eq:GlmbDensity}
  \mathring{f} (\mathring{X})
  =
  \Delta (\mathring{X})
  \sum\limits_{\xi \in \Xi}
  w^{(\xi)} (\mathcal{L} (\mathring{X}))
  [p^{(\xi)}]^{\mathring{X}} \,,
\end{align}
where each $\xi\in\Xi$ represents a history of measurement association maps, each $p^{(\xi)}(\cdot,\ell)$ is a probability density on $\mathbb{X}$, and each weight $w^{(\xi)}$ is non-negative with ${\sum\limits_{(I,\xi)\in \mathcal{F}(\mathbb{L})\times\Xi} w^{(\xi)}(I)=1}$.  The label of a labeled state $\mathring{x}$ is recovered by $\mathcal{L}(\mathring{x})$, where $\mathcal{L} : \mathbb{X} \times \mathbb{L} \mapsto \mathbb{L}$ is the projection defined by $\mathcal{L}((\boldsymbol{x}, \ell)) \triangleq \ell$.  Similarly, for LRFSs, $\mathcal{L}(\mathring{X}) \triangleq \{\mathcal{L}(\mathring{x}) : \mathring{x} \in \mathring{X}\}$.  The distinct label indicator  $\Delta(\mathring{X})=\delta_{(|\mathring{X}|)}(|\mathcal{L}(\mathring{X})|)$ ensures that only sets with distinct labels are considered.

\begin{proposition}
  Given a GLMB density $\mathring{f} (\mathring{X})$ of the form of Equation~(\ref{eq:GlmbDensity}), the cardinality of objects inside a bounded FoV $\mathcal{S}$ is distributed according to
\begin{align}
  \label{eq:GlmbFovCardinality}
  \rho_{\mathcal{S}}(n)
  =&
  \sum\limits_{\mathclap{(\xi, \mathcal{I}_{1} \uplus \mathcal{I}_{2}) \in \Xi \times \mathcal{F}(\mathbb{L})}}
  w^{(\xi)}(I)
  \delta_{n}(|\mathcal{I}_{1}|)
  \left<1_{\mathcal{S}}, p\right>^{\mathcal{I}_{1}}
  \left<1-1_{\mathcal{S}}, p\right> ^{\mathcal{I}_{2}}
\end{align}
\end{proposition}

\textit{Proof}:
Equation~(\ref{eq:ConditionalCardinality}) can be rewritten to accommodate the labeled RFS as
\begin{align}
  \rho_{\mathcal{S}}(n \,|\, \mathring{X}) = \sum_{\mathring{X}^{n} \subseteq \mathring{X}} [1_{\mathcal{S}}(\cdot)]^{\mathring{X}^{n}} [1 - 1_{\mathcal{S}}(\cdot)]^{\mathring{X}\setminus \mathring{X}^{n}}
  \label{eq:ConditionalCardinalityLabeled}
\end{align}
If $\mathring{X}$ is distributed according to the LRFS density $\mathring{f}(\mathring{X})$, the FoV cardinality distribution is obtained via the set integral
\begin{align*}
  \rho_{\mathcal{S}}(n) = \int \rho_{\mathcal{S}}(n\,|\,\mathring{X}) \mathring{f}(\mathring{X}) \delta \mathring{X}
\end{align*}
Expanding the integral,
\begin{align*}
  &\rho_{\mathcal{S}}(n) \nonumber \\
  &=
  \sum_{m=n}^{\infty} \frac{1}{m!}
  \sum_{(\ell_{1} \varlist \ell_{m}) \in
  \mathbb{L}^{m}} \,
  \int\limits_{\mathclap{\mathbb{X}^{m}}}
  \rho_{\mathcal{S}}(n\,|\,\{(\boldsymbol{x}_{1},\ell_{1}) \varlist (\boldsymbol{x}_{m},\ell_{m})\}) \nonumber \\
  &\qquad\cdot
  \mathring{f}(\{(\boldsymbol{x}_{1},\ell_{1}) \varlist (\boldsymbol{x}_{m},\ell_{m})\}) \mathrm{d} \boldsymbol{x}_{1} \cdots \mathrm{d} \boldsymbol{x}_{m}
\end{align*}
Defining $p^{(\xi,\ell)}(x)\triangleq p^{(\xi)}(x,\ell)$, substitution of Equations~(\ref{eq:GlmbDensity}) and~(\ref{eq:ConditionalCardinalityLabeled}) yields
\begin{align*}
  \rho_{\mathcal{S}}(n)
  &=
  \sum_{m=n}^{\infty} \frac{1}{m!}
  m!
  \sum_{\{\ell_{1},\dots,\ell_{m}\} \in \mathbb{L}^{m}}
  \sum_{\xi \in \Xi}
  w^{(\xi)}(\{\ell_{1}, \dots, \ell_{m}\}) \nonumber \\
  &\qquad
  \sum_{\mathclap{I^{n} \subseteq \{\ell_{1}, \dots \ell_{m}\}}}
  \quad
  \big<1_{\mathcal{S}},p^{(\xi,\cdot)}\big>^{I^{n}}
  \big<1 - 1_{\mathcal{S}},p^{(\xi,\cdot)}\big>^{\{\ell_{1}, \dots, \ell_{m}\} \setminus I^{n}} \nonumber \\[1em]
  &=
  \quad
  \sum_{\mathclap{(\xi, I) \in \Xi \times \mathcal{F}(\mathbb{L})}}
  w^{(\xi)}(I)
  \sum_{I^{n} \subseteq I}
  \big<1_{\mathcal{S}},p^{(\xi,\cdot)}\big>^{I^{n}}
  \big<1 - 1_{\mathcal{S}},p^{(\xi,\cdot)}\big>^{I \setminus I^{n}}
\end{align*}
from which Equation~(\ref{eq:GlmbFovCardinality}) follows.
\null\hfill$\square$

\textit{Remark}: Substitution of $n=0$ in Equation \ref{eq:GlmbFovCardinality} gives the GLMB void probability functional \cite[Eq. 22]{BeardVoidProbabilitiesCauchySchwarzGlmb17}, which, while less general, has theoretical significance and practical applications in sensor management.
\section{Sensor Placement Example}
\label{sec:SensorPlacementExample}
\begin{figure}[htbp]
  \centering
  \includegraphics[width=\linewidth]{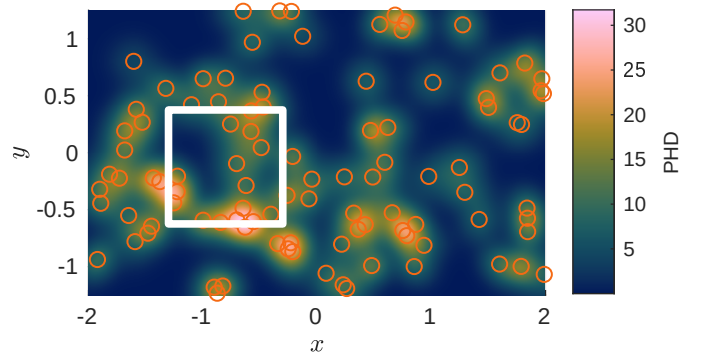}
  \caption{PHD of MB workspace distribution with 100 potential objects, where object means are represented by orange circles and the bounds of the FoV that maximizes the FoV cardinality variance are shown in white.}
  \label{fig:ExampleMaxCardVarPhd}
\end{figure}

The FoV statistics developed in this paper are demonstrated through a sensor placement optimization problem subject to multi-object uncertainty.  For brevity, the workspace distribution is assumed MB-distributed. Numerical simulation is performed for the case of $100$ MB components, with probabilities of existence randomly chosen between $0.35$ and $1$.  Each MB component has a Gaussian density and randomly chosen mean and covariance.  To visualize the workspace distribution, the PHD is shown in Figure~\ref{fig:ExampleMaxCardVarPhd}.

\begin{figure}[htbp]
  \centering
  \includegraphics[width=\linewidth]{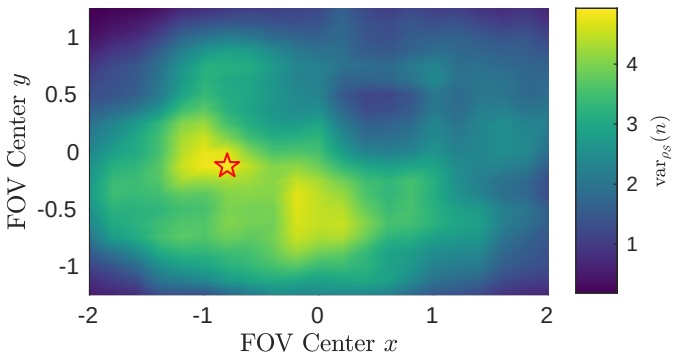}
  \caption{FoV cardinality variance as a function of FoV center location, where the red star denotes the maximum variance point.}
  \label{fig:ExampleMaxCardVarVar}
\end{figure}

The objective of the sensor control problem is to place the FoV, comprised of a square of $1\times 1$ dimensions, in the ROI (Figure \ref{fig:ExampleMaxCardVarPhd}) such that the variance of object cardinality inside the FoV is maximized.  This objective can be interpreted as placing the FoV in a region of the workspace where the object cardinality is most uncertain.  A related objective which minimizes the variance of the \textit{global} cardinality using CB-MeMBer predictions was first proposed in \cite{HoangSensorManagementMultiBernoulliRenyiMaxCardinalityVariance14}. For each candidate FoV placement, the FoV cardinality pmf is given by Equation~(\ref{eq:MbFovCardinality}) and is efficiently approximated using Algorithm~\ref{alg:StochasticMbCardinality}.  The variance of the resulting pmf is shown as a function of the FoV center location in Figure~\ref{fig:ExampleMaxCardVarVar}.  The optimal FoV center location is found to be $(-0.8, -1.25)$.

A compelling result is that, by virtue of the bounded FoV geometry, spatial information is encoded in the FoV cardinality pmf.  It can be seen that the optimal FoV (Fig.~\ref{fig:ExampleMaxCardVarPhd}) has boundary segments (lower half of left boundary and right half of lower boundary) that bisect clusters of MB components.  These boundary segments divide the components' single-object densities such that significant mass appears inside and outside the FoV, increasing the overall FoV cardinality variance.

\section{Conclusions}
\label{sec:Conclusions}
This paper presents an approach for incorporating bounded field-of-view (FoV) geometry into state density updates and object cardinality predictions via finite set statistics.  Negative information is processed in state density updates via a novel Gaussian splitting algorithm that recursively refines a Gaussian mixture approximation near the boundaries of the discrete FoV geometry.  Using FISST, cardinality probability mass functions that describe the probability that a given number of targets exist inside the FoV are derived.  The approach is presented for representative labeled random finite set  distributions and, thus, is applicable to a wide range of tracking, perception, and sensor planning problems.
\bibliographystyle{IEEEtran}
\bibliography{keithlegrand_refs}

\begin{thebibliography}{10}
\providecommand{\url}[1]{#1}
\csname url@samestyle\endcsname
\providecommand{\newblock}{\relax}
\providecommand{\bibinfo}[2]{#2}
\providecommand{\BIBentrySTDinterwordspacing}{\spaceskip=0pt\relax}
\providecommand{\BIBentryALTinterwordstretchfactor}{4}
\providecommand{\BIBentryALTinterwordspacing}{\spaceskip=\fontdimen2\font plus
\BIBentryALTinterwordstretchfactor\fontdimen3\font minus
  \fontdimen4\font\relax}
\providecommand{\BIBforeignlanguage}[2]{{%
\expandafter\ifx\csname l@#1\endcsname\relax
\typeout{** WARNING: IEEEtran.bst: No hyphenation pattern has been}%
\typeout{** loaded for the language `#1'. Using the pattern for}%
\typeout{** the default language instead.}%
\else
\language=\csname l@#1\endcsname
\fi
#2}}
\providecommand{\BIBdecl}{\relax}
\BIBdecl

\bibitem{MahlerStatisticalMultitargetFusion07}
R.~P. Mahler, \emph{Statistical Multisource-Multitarget Information
  Fusion}.\hskip 1em plus 0.5em minus 0.4em\relax Artech House Boston, 2007.

\bibitem{VoLabeledRfsGlmbFilter14}
B.-N. Vo, B.-T. Vo, and D.~Phung, ``Labeled random finite sets and the {B}ayes
  multi-target tracking filter,'' \emph{IEEE Transactions on Signal
  Processing}, vol.~62, no.~24, pp. 6554--6567, 2014.

\bibitem{ReuterLmbFilter14}
S.~Reuter, B.~T. Vo, B.~N. Vo, and K.~Dietmayer, ``The labeled
  multi-{B}ernoulli filter,'' \emph{IEEE Transactions on Signal Processing},
  vol.~62, no.~12, pp. 3246--3260, 2014.

\bibitem{Garcia-fernandezGaussianImplementationPmbm19}
{\'{A}}.~F. Garc{\'{i}}a-Fern{\'{a}}ndez, Y.~Xia, K.~Granstr{\"{o}}m,
  L.~Svensson, and J.~L. Williams, ``{G}aussian implementation of the
  multi-{B}ernoulli mixture filter,'' in \emph{2019 22nd International
  Conference on Information Fusion (FUSION)}, 2019.

\bibitem{HoangSensorManagementMultiBernoulliRenyiMaxCardinalityVariance14}
\BIBentryALTinterwordspacing
H.~G. Hoang and B.~T. Vo, ``Sensor management for multi-target tracking via
  multi-{B}ernoulli filtering,'' \emph{Automatica}, vol.~50, no.~4, pp.
  1135--1142, 2014. [Online]. Available:
  \url{http://dx.doi.org/10.1016/j.automatica.2014.02.007}
\BIBentrySTDinterwordspacing

\bibitem{BeardVoidProbabilitiesCauchySchwarzGlmb17}
M.~Beard, B.~T. Vo, B.~N. Vo, and S.~Arulampalam, ``Void probabilities and
  {C}auchy-{S}chwarz divergence for generalized labeled multi-{B}ernoulli
  models,'' \emph{IEEE Transactions on Signal Processing}, vol.~65, no.~19, pp.
  5047--5061, 2017.

\bibitem{WangMultiSensorControlLmb18}
\BIBentryALTinterwordspacing
X.~Wang, R.~Hoseinnezhad, A.~K. Gostar, T.~Rathnayake, B.~Xu, and
  A.~Bab-hadiashar, ``Multi-sensor control for multi-object {B}ayes filters,''
  \emph{Signal Processing}, vol. 142, pp. 260--270, 2018. [Online]. Available:
  \url{http://dx.doi.org/10.1016/j.sigpro.2017.07.031}
\BIBentrySTDinterwordspacing

\bibitem{FerrariGeometricDetectIntercept09}
\BIBentryALTinterwordspacing
S.~Ferrari, R.~Fierro, B.~Perteet, C.~Cai, and K.~Baumgartner, ``A geometric
  optimization approach to detecting and intercepting dynamic targets using a
  mobile sensor network,'' \emph{SIAM Journal on Control and Optimization},
  vol.~48, no.~1, pp. 292--320, January 2009. [Online]. Available:
  \url{https://doi.org/10.1137/07067934X}
\BIBentrySTDinterwordspacing

\bibitem{WeiGeometricTransversalsSensorPlanning15}
H.~Wei and S.~Ferrari, ``A geometric transversals approach to sensor motion
  planning for tracking maneuvering targets,'' \emph{IEEE Transactions on
  Automatic Control}, vol.~60, no.~10, pp. 2773--2778, 2015.

\bibitem{GehlySearchDetectTrack18}
S.~Gehly, B.~A. Jones, and P.~Axelrad, ``Search-detect-track sensor allocation
  for geosynchronous space objects,'' \emph{IEEE Transactions on Aerospace and
  Electronic Systems}, vol.~54, no.~6, pp. 2788--2808, 2018.

\bibitem{BuonviriSurveyMultiSensorRfs19}
A.~Buonviri, M.~York, K.~A. LeGrand, and J.~Meub, ``Survey of challenges in
  labeled random finite set based distributed multi-sensor multi-object
  tracking,'' in \emph{2019 IEEE Aerospace Conference}, 2019.

\bibitem{KochExploitingNegativeEvidence07}
W.~Koch, ``On exploiting `negative' sensor evidence for target tracking and
  sensor data fusion,'' \emph{Information Fusion}, 2007.

\bibitem{SongTargetTrackingStateDependentPd11}
T.~L. Song, D.~Musicki, and K.~D. Sol, ``Target tracking with target state
  dependent detection,'' \emph{IEEE Transactions on Signal Processing},
  vol.~59, no.~3, pp. 1063--1074, 2011.

\bibitem{WeiDistributedSpaceTrackingFovEm18}
B.~Wei and B.~Nener, ``Distributed space debris tracking with consensus labeled
  random finite set filtering,'' \emph{Sensors}, vol.~18, pp. 1--26, 2018.

\bibitem{BaumgartnerOptimalControlUnderwaterSensor09}
K.~A.~C. Baumgartner, S.~Ferrari, and A.~V. Rao, ``Optimal control of an
  underwater sensor network for cooperative target tracking,'' \emph{IEEE
  Journal of Oceanic Engineering}, vol.~34, no.~4, pp. 678--697, 2009.

\bibitem{VoGaussianMixturePhd06}
B.-N. Vo and W.-K. Ma, ``The {G}aussian mixture probability hypothesis density
  filter,'' \emph{IEEE Transactions on Signal Processing}, vol.~54, no.~11, pp.
  4091--4104, Nov 2006.

\bibitem{LegrandRelativeSpaceObjectTrackingFusion15}
K.~A. LeGrand and K.~J. DeMars, ``Relative multiple space object tracking using
  intensity filters,'' in \emph{18th International Conference on Information
  Fusion (FUSION)}.\hskip 1em plus 0.5em minus 0.4em\relax IEEE, 2015, pp.
  1253--1261.

\bibitem{HuberEntropyApproximationSplitLibrary08}
M.~F. Huber, T.~Bailey, H.~Durrant-Whyte, and U.~D. Hanebeck, ``On entropy
  approximation for {G}aussian mixture random vectors,'' \emph{IEEE
  International Conference on Multisensor Fusion and Integration for
  Intelligent Systems}, pp. 181--188, 2008.

\bibitem{DemarsEntropyPropagationGaussSplit13}
\BIBentryALTinterwordspacing
K.~J. DeMars, R.~H. Bishop, and M.~K. Jah, ``Entropy-based approach for
  uncertainty propagation of nonlinear dynamical systems,'' \emph{Journal of
  Guidance, Control, and Dynamics}, vol.~36, no.~4, pp. 1047--1057, 2013.
  [Online]. Available: \url{http://arc.aiaa.org/doi/10.2514/1.58987}
\BIBentrySTDinterwordspacing

\bibitem{MahlerAdvancesStatisticalMultitargetFusion14}
R.~P. Mahler, \emph{Advances in Statistical Multisource-Multitarget Information
  Fusion}.\hskip 1em plus 0.5em minus 0.4em\relax Artech House, 2014.

\bibitem{WilliamsMarginalMultiBernoulliMixturePGFL15}
J.~L. Williams, ``Marginal multi-{B}ernoulli filters: {RFS} derivation of
  {MHT}, {JIPDA}, and association-based {MeMBer},'' \emph{IEEE Transactions on
  Aerospace and Electronic Systems}, vol.~51, no.~3, pp. 1664--1687, 2015.

\bibitem{VoLabeledRfsConjugatePriors13}
B.-T. Vo and B.-N. Vo, ``Labeled random finite sets and multi-object conjugate
  priors,'' \emph{IEEE Transactions on Signal Processing}, vol.~61, no.~13, pp.
  3460--3475, 2013.

\end{thebibliography}

\appendices
\section{Proof of Proposition~\ref{pro:MbFovCardinality}}
\label{a:ProofOfMbFovCardinality}
Equation~(\ref{eq:MbDensity}) can be rewritten as
\begin{align}
  f(X)
  =
  \left[
    \left(
      1 - r^{(\cdot)}
    \right)
  \right]^{\{1..M\}}
  \sum\limits_{(\mathcal{I}_{\sigma}) \uplus \mathcal{I}_{3}}
  \left[
    \dfrac{%
      r^{i_{(\cdot)}} p^{i_{(\cdot)}}(x_{(\cdot)})
    }{%
      1 - r^{i_{(\cdot)}}
    }
  \right]^{\{1..n\}}
  \label{eq:MbDensityIndexSet}
\end{align}
where $(\mathcal{I}_{\sigma})$ denotes the (ordered) sequence $(i_{1} \varlist i_{n}) = (\alpha_{\sigma(1)}\varlist \alpha_{\sigma(n)})$, where the $n$-tuple index set $\{\alpha_{1} \varlist \alpha_{n}\} \uplus \mathcal{I}_{3}=\{1\varlist M\}$ and $\sigma$ is a permutation of $\{1\varlist n\}$.

Substituting Equation~(\ref{eq:MbDensityIndexSet}) into Equation~(\ref{eq:FovCardinalityGeneralExpanded}),
\begin{align*}
  &\rho_{\mathcal{S}}(n)
  =
  \left[
    \left(
      1 - r^{(\cdot)}
    \right)
  \right]^{\{1.. M\}} \nonumber \\
  &\quad
  \cdot
  \sum_{m=n}^{M}
  \frac{1}{m!}
  \int_{\mathbb{X}^{m}}
  \sum\limits_{(\mathcal{I}_{\sigma}) \uplus \mathcal{I}_{3}}
  \delta_{m}(|\mathcal{I}_{\sigma}|)
  \left[
    \dfrac{%
      r^{i_{(\cdot)}} p^{i_{(\cdot)}}(\boldsymbol{x}_{(\cdot)})
    }{%
      1 - r^{i_{(\cdot)}}
    }
  \right]^{\{1..m\}}\nonumber\\
  &\qquad
  \sum_{X^{n}\subseteq X} [1_{\mathcal{S}}(\cdot)]^{X^{n}} [1 - 1_{\mathcal{S}}(\cdot)]^{X \setminus X^{n}}
  \mathrm{d} \boldsymbol{x}_{1} \cdots \mathrm{d} \boldsymbol{x}_{m}
\end{align*}
The last sum can be written in terms of label index sets $\mathcal{I}_{1}\uplus \mathcal{I}_{2}=\mathcal{I}_{\sigma}$ as
\begin{align*}
  &\rho_{\mathcal{S}}(n)
  =
  \left[
    \left(
      1 - r^{(\cdot)}
    \right)
  \right]^{\{1..M\}} \nonumber \\
  &\quad
  \cdot
  \sum_{m=n}^{M}
  \frac{1}{m!}
  \int_{\mathbb{X}^{m}}
  \sum\limits_{(\mathcal{I}_{\sigma}) \uplus \mathcal{I}_{3}}
  \delta_{m}(|\mathcal{I}_{\sigma}|)
  \left[
    \dfrac{%
      r^{i_{(\cdot)}} p^{i_{(\cdot)}}(\boldsymbol{x}_{(\cdot)})
    }{%
      1 - r^{i_{(\cdot)}}
    }
  \right]^{\{1..m\}}\nonumber\\
  &\quad
  \cdot
  \sum_{\mathclap{\mathcal{I}_{1} \uplus \mathcal{I}_{2}=\mathcal{I}_{\sigma}}}
  \delta_{n}(|\mathcal{I}_{1}|)
  [1_{\mathcal{S}}(\boldsymbol{x}_{(\cdot)})]^{\{j:i_{j}\in\mathcal{I}_{1}\}}
  [1 - 1_{\mathcal{S}}(\boldsymbol{x}_{(\cdot)})]^{\{j:i_{j}\in \mathcal{I}_{2}\}} \nonumber\\
  &\qquad \,
  \mathrm{d} \boldsymbol{x}_{1} \cdots \mathrm{d} \boldsymbol{x}_{m}
\end{align*}
Distributing terms from the second summation,
\begin{align*}
  &\rho_{\mathcal{S}}(n)
  =
  \left[
    \left(
      1 - r^{(\cdot)}
    \right)
  \right]^{\{1..M\}} \nonumber\\
  &\quad
  \cdot
  \sum_{m=n}^{M}
  \frac{1}{m!}
  \int_{\mathbb{X}^{m}}
  \sum\limits_{(\mathcal{I}_{\sigma}) \uplus \mathcal{I}_{3}}
  \delta_{m}(|\mathcal{I}_{\sigma}|)
  \sum_{\mathcal{I}_{1} \uplus \mathcal{I}_{2}=\mathcal{I}_{\sigma}} \delta_{n}(|\mathcal{I}_{1}|) \nonumber\\
  &\quad
  \cdot
  \left[
    \dfrac{%
    1_{\mathcal{S}}(\boldsymbol{x}_{(\cdot)})r^{i_{(\cdot)}} p^{i_{(\cdot)}}(\boldsymbol{x}_{(\cdot)})
    }{%
      1 - r^{i_{(\cdot)}}
    }
  \right]^{\{j:i_{j}\in\mathcal{I}_{1}\}} \nonumber \\
  &\quad
  \cdot
  \left[
    \dfrac{%
      [1 - 1_{\mathcal{S}}(\boldsymbol{x}_{(\cdot)})]
      r^{i_{(\cdot)}} p^{i_{(\cdot)}}(\boldsymbol{x}_{(\cdot)})
    }{%
      1 - r^{i_{(\cdot)}}
    }
  \right]^{\{j:i_{j}\in \mathcal{I}_{2}\}}
  \mathrm{d} \boldsymbol{x}_{1} \cdots \mathrm{d} \boldsymbol{x}_{m}
\end{align*}
Because $\mathcal{I}_{1} \cap \mathcal{I}_{2}= \emptyset$, then $\{\boldsymbol{x}_{j} : i_{j}\in \mathcal{I}_{1} \} \cap \{ \boldsymbol{x}_{j} : i_{j}\in \mathcal{I}_{2} \} = \emptyset $ and the integral on $\mathbb{X}^{m}$ becomes a product of integrals on $\mathbb{X}$, such that
\begin{align*}
  &\rho_{\mathcal{S}}(n)
  =
  \left[
    \left(
      1 - r^{(\cdot)}
    \right)
  \right]^{\{1..M\}} \nonumber \\
  &\quad
  \cdot
  \sum_{m=n}^{M}
  \frac{1}{m!}
  \sum\limits_{(\mathcal{I}_{\sigma}) \uplus \mathcal{I}_{3}}
  \delta_{m}(|\mathcal{I}_{\sigma}|)
  \sum_{\mathcal{I}_{1} \uplus \mathcal{I}_{2}=\mathcal{I}_{\sigma}} \delta_{n}(|\mathcal{I}_{1}|) \nonumber\\
  \qquad &
  \cdot
  \left[
    \dfrac{%
      \left< 1_{\mathcal{S}}, r^{i_{(\cdot)}} p^{i_{(\cdot)}}\right>
    }{%
      1 - r^{i_{(\cdot)}}
    }
  \right]^{\{j:i_{j}\in\mathcal{I}_{1}\}}
  \left[
    \dfrac{%
      \left<1 - 1_{\mathcal{S}},
      r^{i_{(\cdot)}} p^{i_{(\cdot)}}\right>
    }{%
      1 - r^{i_{(\cdot)}}
    }
  \right]^{\{j:i_{j}\in \mathcal{I}_{2}\}}
\end{align*}
Now note that the result of the innermost sum does not depend the permutation order of $(\mathcal{I}_{\sigma})$.  Thus the property \cite[Lemma 12]{VoLabeledRfsConjugatePriors13} that for an arbitrary symmetric function $h$
\begin{align*}
  \sum_{(i_{1}, \ldots, i_{m})} h(\{i_{1}, \ldots, i_{m}\})
  = m!
  \sum_{\{i_{1}, \ldots, i_{m}\}} h(\{i_{1}, \ldots, i_{m}\})
\end{align*}
is applied, yielding
\begin{align*}
  &\rho_{\mathcal{S}}(n)
  =
  \left[
    \left(
      1 - r^{(\cdot)}
    \right)
  \right]^{\{1, \hdots, M\}} \nonumber \\
  &\quad
  \cdot
  \sum_{m=n}^{M}
  \sum\limits_{\mathcal{I}_{1} \uplus \mathcal{I}_{2}\uplus \mathcal{I}_{3}}
  \delta_{m}(|\mathcal{I}_{1}\uplus \mathcal{I}_{2}|)
  \delta_{n}(|\mathcal{I}_{1}|) \nonumber \\
  &\quad
  \cdot
  \left[
    \dfrac{%
      \left< 1_{\mathcal{S}}, r^{(\cdot)} p^{(\cdot)}\right>
    }{%
      1 - r^{(\cdot)}
    }
  \right]^{\mathcal{I}_{1}}
  \left[
    \dfrac{%
      \left<1 - 1_{\mathcal{S}},
      r^{(\cdot)} p^{(\cdot)}\right>
    }{%
      1 - r^{(\cdot)}
    }
  \right]^{\mathcal{I}_{2}}
\end{align*}
The term $\delta_{m}(|\mathcal{I}_{1} \uplus \mathcal{I}_{2}|)$ is non-zero only when the combined cardinality of $\mathcal{I}_{1}$ and $\mathcal{I}_{2}$ is equal to $m$---the index of the outermost sum.  Thus, the outermost sum is absorbed by the second sum to give Equation~(\ref{eq:MbFovCardinality}).
\null\hfill$\square$

\end{document}